# Highlighting Ukrainian Contributions to the Development of the Field of Particle Physics

**T. Hryn'ova (CNRS, France), M. Strikha (Taras Shevchenko National University of Kyiv, Ukraine), V. Grekov (Departmental Science and Technology Museum "UIPT Museum and Cultural Complex" of the National Science Center "Kharkiv Institute of Physics and Technology" of National Academy of Sciences of Ukraine, Kharkiv, Ukraine)**

## 1. Introduction

Ukrainian scientific contributions have often been attributed to the "Russian Empire" or the "Soviet Union" (USSR), as well as to the multitude of countries to which scientists born on the territory of Ukraine emigrated to save their lives and to continue their research. Although both the Russian Empire and the USSR were multiethnic states, in English all their inhabitants are frequently misleadingly referred to as "Russian"[1]. The major challenge of this paper is thus to define who are *Ukrainian* scientists, prior to the declaration of independence of Ukraine in 1991 and determine their contributions in the field of particle physics.

Although the early roots of the Ukraine national idea can be traced back to the Polish-Lithanian Commonwealth (1569-1795) [Plokhy, 2010; Snyder 2003], most of its modern development took part under the empires, but for a brief period of independent states in 1917-1921 following the aftermath of the World War I (WWI). From 1795 to 1917, the territory of the present-day Ukraine was split first between the Russian Empire/the Austrian Empire[2] and, in 1922-1939, the USSR/Poland, with small parts in Romania and Czechoslovakia.

The political situation for Ukrainians within all these states was essentially different: in the Russian Empire till 1905 there were no political freedoms and the Ukrainian language was totally forbidden/severely restricted in public usage; while the Austrian Empire since 1848 was a parliamentary monarchy with Ukrainian language partly accepted in the local administrations, education, etc. Therefore part of West Ukraine under Habsburg dynasty rule was treated in the beginning of the XX century as a kind of "Ukrainian Piemont", a stronghold for the national institutions.

Ukraine's internationally recognized borders took their current shape in 1954, when it was the Ukrainian Soviet Socialist Republic (Ukrainian SSR), a constituent republic of the USSR. However, the Ukrainian SSR had a formal status of a "sovereign contractual state" within the USSR[3], and was a founding member of the United Nations organization in 1945 (separate from the USSR).

---

[1] From Oxford English Dictionary [Oxford]: "Russian 1.a. A member of an East Slavonic-speaking people living primarily in Russia and neighbouring countries; a native or inhabitant of Russia, or (more widely) its former empire or the former Soviet Union; a person of Russian nationality."

[2] From 1867 - the Austro-Hungarian Empire.

[3] "суверенна договірна держава" [Constitution UkrSSR, 1929, Art. I.3]

The period of establishment of the Ukrainian nation coincides with development of systematic understanding of the building blocks of our universe. From the introduction of the concept of “a chemical element” [Lavoisier, 1789], through the discovery of the structure of an atom [Thomson, 1897; Rutherford, 1911], followed by the emergence of the quantum theory in 1900s - 1930s, discoveries of the zoo of subatomic particles in 1930s - 1950s, development of the quark model and understanding of the fundamental forces in 1960s - 1970s and finally the establishment and validation of the Standard model of particle physics from 1970s to the present day. Scientists born in Ukraine, working in the fields of chemistry, physics, mathematics and engineering played an important role in these developments.

This work is meant as a step towards restoration of historical justice, returning from oblivion the forgotten names and reconstructing the Ukrainian contributions to the development of particle physics. Section 2 reviews leading Ukrainian institutions in the field of physics and technology. Section 3 discusses the working environment in these institutes prior to independence of Ukraine. Section 4 explains the role of Ukrainian language as a marker of Ukrainian national identity. Section 5 summarizes the impact of Russification on the population of Ukraine, while Section 6 chronicles waves of emigration and their impact on the Ukrainian scientific diaspora. Section 7 outlines criteria used in this work to define scientists as “Ukrainian”. Section 8 talks about first physics papers in Ukrainian as well as research which led to development of particle physics. Sections 9-11 present Ukrainian contributions to particle physics theory, accelerators and experiments, respectively. Section 12 tells about the scientific diplomacy of scholars from Ukraine.

## 2. Leading institutions in Ukraine in the field of physics and technology

The first University on the territory of Ukraine was established[4] in 1784 in Lviv[5], which was then part of the Austrian Empire. The first university on Ukrainian territories which were then part of the Russian Empire was founded in Kharkiv[6] in 1804/1805. Universities were also established in Kyiv[7] in 1834, Odesa in 1865 and Chernivtsi in 1875. Although these universities operated entirely under imperial control and Ukrainian was not the language of instruction, they played a key role in the formation of Ukrainian intelligentsia. The university scholars and students initiated studies of Ukrainian history, language, and culture, leading to Ukrainian national revival [Snyder, 2003]. Three more universities, in Dnipro[8], in Kamjanets-Podilskyj and in Simferopol[9], were founded in 1918 during Ukraine’s struggle for independence. Universities in Soviet Ukraine were dissolved between 1920 and 1933 and, during this period, were reorganized into Institutes of Public Education, which led to a decline in academic standards and the departure of many students and academic personnel. Universities in Uzhgorod and Donetsk were established in 1945 and 1965, respectively.

[4] “The university's precursor was Lviv's Jesuit Academy (referred to in some internal documents as 'university') that operated in the city between 1661 and 1773”, when it was a part of the Polish–Lithuanian Commonwealth. [Bilenky]

[5] In 1772, following the first partition of Poland, it became part of Austrian-Hungarian Empire till its disintegration in 1918, when the city briefly became the capital of the Western Ukrainian People’s Republic. It was part of the Polish Republic in 1918-1939.

[6] Kharkiv was the capital of the Ukrainian SSR in 1919-1934.

[7] Kyiv was the capital of the Ukrainian SSR in 1934-1991 and remained the capital of Ukraine after its independence.

[8] Dnipro was called Ekaterinoslav/Yekaterinoslav till 1926 (but a short period of Ukrainian independence in 1919, when it was Sicheslav), and Dnipropetrovsk till 2016.

[9] It was formed originally as the Tavria branch of the University of Kyiv.

In the 19th century, when industrialization increased the demand for engineers, scientists, and technical specialists across the empires, a number of polytechnical institutes were established: the first one was in Lviv in 1877, the institutes in Kharkiv in 1885, in Kyiv 1898, in Dnipro 1899 and in Odesa in 1918 were founded in close succession.

In parallel with the higher education institutions mentioned above, a number of dedicated research institutes focusing on both fundamental physics and materials science were established in two waves, in the 1920s and the 1960s. The second wave often expanded on branches of earlier institutions. All of these institutes were founded within the framework of the National Academy of Sciences of Ukraine, which was established in 1918 as the country's principal scientific organization coordinating research across disciplines. A selected list of these establishments, most relevant to research in particle physics, is provided below:

- **Institute of Physics** (1926, Kyiv) - IoP initially focused on solid-state physics, semiconductors, dielectrics, and later nuclear and particle physics as well.
- **Ukrainian Institute of Physics and Technology** (1928, Kharkiv) - UIPT was the pioneering center for nuclear and particle physics in the Soviet Union, founded in Kharkiv, then capital of the Ukrainian SSR. Now known as **the National Scientific Center "Kharkiv Institute of Physics and Technology" (KIPT)**. See more in [Ranyuk, 2006].
- **Institute for Single Crystals (1955, Kharkiv)** – ISC specializes in the growth and study of single crystals, with a focus on optical materials, scintillators, and functional crystals for scientific and industrial applications. See more in [Tarasov, 2026; Tupitsyna, 2026].
- **V. Lashkaryov Institute of Semiconductor Physics** (1960) – a key institution in semiconductor research and microelectronics.
- **Donetsk Institute for Physics and Engineering named after O.O. Galkin** (1965-2014 in Donetsk, now in Kyiv) – focused on research in solid-state physics and materials science.
- **Bogolyubov Institute for Theoretical Physics** (1966, Kyiv) – BITP is a leading center for theoretical physics. See more in [Gorenstein, 2026].
- **Institute for Nuclear Research** (1970, Kyiv) – KNIR focuses on nuclear physics, nuclear energy, and related technologies.
- **Institute of Applied Physics** (1991, Sumy) – IAP specializes in applied physics, including plasma physics and electronics.
- **Institute for Scintillation Materials (2002, Kharkiv)** – ISMA was established on the basis of the Institute for Single Crystals and focuses on the development of scintillation materials and detectors for medical imaging and high-energy physics.

The institutional formation of physical science in Ukraine evolved from the classical universities of the 18th–19th centuries to the creation of specialized research centers. KIPT has played a pioneering role in the development of particle physics in Ukraine since its establishment in 1928. Polytechnic institutes met the demand for engineering personnel, while two waves of the creation of academic institutions (in the 1920s and 1960s) formed the foundation for particle physics research in Ukraine. This extensive network of scientific institutions and universities not only accumulated significant intellectual potential and attracted talent from other parts of the former USSR[10], but also led integration of Ukraine into global projects in high-energy physics.

### 3. Working environment in the Ukrainian scientific institutions prior to independence

It is often overlooked that in late 19th and early 20th century (during the last decades of the Russian Empire and the first two decades of the Soviet Union) a significant fraction of scientists from Ukraine was able to benefit from education in Western Europe as well as extensive research trips abroad [Timoshenko, 1968; Gamow, 1970; Bogolyubov, 1996]. Some of these scientists, like Kyrylo Synelnykov, came back to Ukraine with their foreign wives [Lucie Str., 1944]. UIPT had not only foreign visitors, including Bohr, Van der Graaff, Dirac [Tan'shyna, 2019], but also foreign researchers on its staff. In "*My Life as Physicist*" Victor Weisskopf recalls: "In 1933 I went to Kharkov, Russia[11] for almost a year, where it was possible to get a job with subsistence. Working in Kharkov at that time were Landau, Lifschitz, and Achiezer, and many other young Russian physicists[12]. Life in Russia was by no means easy. There was a serious famine[13] because of the difficulties of collectivization. But life was interesting and instructive." [Weisskopf, 1972]. This extract highlights that, in the absence of an independent Ukrainian state at the time, most scientists working in institutions in Ukraine were perceived abroad as "Russian."

The Soviet government grew wary of research freedom and international collaborations of Ukrainian scientists. In the late 1930s, 16 scientists from UIPT were arrested, including Landau, eight of them were subsequently executed [Tan'shyna, 2019], like many other Ukrainian scientists at the time [Timoshyn, 2025]. Foreign scientists were forced to leave the Ukrainian SSR[14], some of them passed through Soviet prisons before that [Weissberg-Cybulski, 1951].

---

[10] Lev Shubnikov (b. Saint Petersburg 1901, UIPT 1930-37), Lev Landau (b. Baku 1908, UIPT 1932-37), Ivan Obreimov (b. Annecy 1894, UIPT 1929-1941), Anton Walter (b. Saint Petersburg 1905, UIPT from 1930), Oleksandr Akhiezer (b. Cherykaw Belarus 1911, KPI 1929-34, UIPT from 1934), Naum Akhiezer (b. Cherykaw Belarus 1901, in Kharkiv from 1922), Oleksandr Leipunsky (b. Drahle Poland 1903, UIPT 1930-1941, IoP in Kyiv 1941-1949), Dmitry Volkov (b. Leningrad 1925, in Kharkiv from 1951), Gennady Zinovjev (b. Birobidzhan 1941, 1958-63 Dnipro University, from 1975 in BITP in Kyiv), Aleksander Kapustnikov (b. Baku 1945, Dnipro University from 1963), etc.

[11] Weisskopf uses russified spelling of Kharkiv, which was prevalent at the time of publication (1970). He also mistakenly located Kharkiv in Russia, while Kharkiv was the capital of Soviet Ukraine.

[12] Weisskopf's statement provides an excellent example of misapplication of the term "Russian physicists". Landau (1908-1968) was born in Baku and studied there till 1924, his parents originally from present-day Belarus; he studied in Leningrad in 1924-1929 and did a research trip abroad in 1929–1931 before coming to work in Kharkiv (1932-1937). Lifschitz was born in Kharkiv. O. Akhiezer was born in present-day Belarus, but worked all his life in Kharkiv.

[13] Weisskopf was in Kharkiv during Holodomor, which resulted in at least 3.3 million deaths in Soviet Ukraine in 1932-1933 [Snyder 2015, p. 53].

[14] The only exception in UIPT was F. Lange (1899-1987), who was not subjected to any repressions [Ranyuk, 2006].

As stated in the report of the first delegation from CERN to the independent Ukraine in 1992 “Indeed the highly centralized systems of the former USSR made it difficult for institutes outside the Moscow or St. Petersburg area to follow an independent policy of scientific contacts. A certain feeling of isolation is the result, despite the fact that Kharkov has produced quite a few eminent physicists (e.g. Landau, Lifshitz, Pomeranchuk). Few people we have met had had the opportunity to go abroad and our visit was for many the first direct contact with CERN.” [CERN, 1992].

Scientific institutions in the Ukrainian SSR indeed were systematically restrained, conservatism and provincialism were imposed on them [Grekov, 2026]. Those involved in secret military projects were not allowed to publish their results. It is to be noted, that before 1991 there always was an essential migration of active scientists[15] from the institutions on Ukrainian territory to imperial centers (Moscow, Leningrad, Novosibirsk, Dubna, etc). Many scientists did not return to Ukraine after WWII. Let us consider the career of Dmytro Ivanenko (1904 - 1994), a theorist born in Poltava, who predicted the proton-neutron structure of the atomic nucleus [Schenderovskyi, 2026]. Ivanenko had studied in Poltava, Kharkiv and in Leningrad, then worked for a short period in UIPT in Kharkiv (1929-1931), moved back to Leningrad, was exiled to Tomsk in Siberia, returned to Ukraine to teach in Kyiv university in 1940-41, and after WWII worked in Moscow. His sister was a popular Ukrainian writer Oksana Ivanenko.

On the occasions when Ukrainian universities were abolished between 1920 and 1933, or when their standards dropped, like after WWII, the best students went to study in Moscow and Leningrad (or Dubna after 1956)[16], joining ranks of their compatriots who moved under the Russian Empire[17]. These centers had a vibrant Ukrainian emigree community [Timoshenko, 1968].

Scientists from Ukraine, who worked or studied in Moscow and Leningrad (or Dubna after 1956), had more opportunities for research trips abroad both in the 1920-30s and after Stalin’s death in 1953. The majority of particle physics research and international collaboration in the Soviet Union was channeled through a new research center created in Dubna, 110 km north of Moscow, in 1956.

In 1975, W.O. Lock published an extensive account of the history of collaboration between CERN and research institutes in the USSR from 1955 to 1970 [Lock, 1975]. The document includes numerous scientists born in Ukraine, such as V.I. Veksler (1907–1966), G.J. Budker (1918–1977) [Johnsen, 1977], A.F. Joffe (1880–1960) [PT, 1961], and V. Yarba (b. 1935), whose work will be discussed further in this paper. It also mentions names of multiple Soviet republics[18]; however, somewhat surprisingly, it contains no mention of “Ukraine” or “Ukrainian,” even though Kyiv—appearing in its Russified form, “Kiev”—is mentioned four times, mostly in its role as the host city of the 1959 International Conference on High-Energy Physics (IX Rochester conference), one of the most prestigious international conferences in particle physics. Although Kyiv was among a few cities world-wide to host this conference twice, in 1959 and 1970 (IX and XV Rochester conferences), it was not associated with Ukraine in the eyes of the international scientific community at the time [Lalli, 2024].

[15] L. Mandelstam, N. Papaleksi and I. Tamm (moved from Odesa in 1922), A. Alexandrov (moved in 1930), D. Ivanenko (moved in 1932), L. Landau (moved in 1937 to avoid purge), E. Lifshitz (moved in 1939 to follow Landau), Bogolyubov (moved 1941, joint posts 1945-1956, 1966-1973), A. Kompaneyets (moved 1941), V. Ogievetsky (moved 1950), V. Baier (moved 1955), I. Khriplovich (moved 1959), I. Lifshitz (joint post 1962-68, moved 1969), M. Azbel (moved 1964) and many others. More discussion in [Azbel, 1981].

[16] Moved to study in Leningrad: G. Gamow (1923), D. Ivanenko (1923), M. Bronstein (1926); to Moscow: G. Budker (1936), I. Khalatnikov (1941) [Teryaeva, 2019], G. Gorelik (graduated in 1972 and remained), to Dubna: L. Jenkovszky (1964-67) [Csörgő, 2025].

[17] Kapitsa family, A. Joffe, S. Timoshenko, I. Tamm.

The Fourth International Symposium on Particle Physics held at CERN in 2025 focused on developments in particle physics from 1980 till 2000. Contributions from “Kiev[19] National University” and the “Kharkov[20] Physical Technical Institute” were mentioned in the presentation “Post 1980s Diaspora of Russian Particle Physicists” [Shiltsev, 2025, p. 7]. It showed that only 3 out of 102 research groups from “Soviet and Russian Labs and Universities” who participated in various Fermilab experiments in 1972-2017 were based in Ukraine. Compared with the fact that in the 1930s about 25% of the research papers on nuclear physics were done in Ukraine [Ranyuk, 2006, p. 265], this shows the devastating impact of the Soviet project on the development of fundamental science research in Ukraine. This talk also highlighted persistent efforts by Russian scientists and diaspora to bring the Ukrainian scientific community into their sphere of influence[21], based on Russian as a common language of communication.

The above highlights that although the Ukrainian SSR (1921-1991) had many formal rights on paper as well as representation in 37 international organizations [Zhuk, 2026], including the United Nations since 1945, its scientific institutions were not recognised as “Ukrainian”.

The working environment in Ukrainian scientific institutes until 1991 was characterized by a contradictory combination of international recognition and imperial constraints. Despite visits by some of the world’s leading scientists (e.g. Niels Bohr, Paul Dirac) and active collaboration with foreign researchers, the Holodomor and Stalin's repressions of the 1930s severely undermined the human resources potential of the UIPT. Further centralization of science in the USSR led to systematic "brain drain" to Moscow and Dubna. Furthermore, the achievements of scientists from Ukraine were often mistakenly labeled as "Russian". Even under conditions of isolation and forced provincialization, Ukrainian centers maintained a high level of research, although their subjectivity in the international arena (in particular, at CERN) remained in the shadow of the Soviet center until the restoration of independence.

[18] “Georgia”, “Georgian”, “Latvia”, “Kazakh Academy of Science”, “Armenian Academy of Science”
[19] Note russified spelling of Kyiv in the original text.
[20] Note the russified spelling of Kharkiv in the original text.
[21] In this effect one should also note the “George Gamow award, established by the Russian-American Association of Scientists (RASA-America, Russian-American Science Association) in memory of the outstanding Russian-American physicist, Professor Georgy Antonovich Gamow (1904-1968) and to encourage members of the Russian-speaking scientific diaspora for outstanding achievements recognized by the wider scientific community” [RASA]. Gamow was born in Odesa in Ukraine in a family connected with Ukrainian national movement, studied/worked in Leningrad in 1922-1933 (of which he spent 1929-1931 abroad and remaining two years trying to escape Soviet Union. Overall he lived 18 years in Odesa, 9 years in Leningrad and 37 years of his life outside the Soviet Union.

## 4. Ukrainian language as a marker of Ukrainian national identity

Prior to Ukraine's declaration of independence in 1991, speaking the Ukrainian language was a key marker of Ukrainian identity. But in the Russian Empire as well as in the USSR, career advancement often required Russian. Stephan Timoshenko (1878-1972), one of the founding members of the Ukrainian Academy of Sciences, remembers in his autobiography [Timoshenko 1968]: "In childhood we spoke a mixture of Russian and Ukrainian, and I grew up without knowing Russian well. That was probably the most troublesome subject for me in the entire realschole curriculum. In speaking Russian I was never able to get rid of my Ukrainian accent. Later, when I entered the Institute of Ways and Communication, the other fellows there called me a khokhol (a derisive nickname meaning Ukrainian)."

While the publications in Ukrainian faced near-total bans up to 1917 in the Russian Empire, scientific papers on physics in Ukrainian had been published in Lviv since 1877 [Strikha, 2024]. The fact that the first such paper by I. Pulyui "About a radiometer or a light mill" was published in Lviv Pravda No. 20 (1877), 772, a "monthly for literature, science, and politics", highlights the importance of physics research for Ukrainian society. The Shevchenko Scientific Society, crucial for the establishment of the Ukrainian scholarly community, was founded shortly after that in Lviv in 1892. The main body of scientific publications in Ukrainian at the time were published in the "Notes of the Shevchenko Scientific Society"[22]. They included both original works of Ukrainian scientists as well as reviews of recent research results published in other languages. The society counted I. Pulyui, S. Timoshenko, A. Joffe, M. Planck, A.Einstein and D. Rozhanskyi among its members [Holovatch, 2018], and is often regarded as an unofficial Ukrainian academy of sciences [Strikha, 2024].

Creation of the Academy of Sciences of Ukraine in 1918 during the short period of Ukrainian independence boosted the use of Ukrainian in its scientific institutions as well as for education. The first specialized physical journal in Ukrainian, Fizychni Zapysky (Physical Notes), was initiated by Oleksandr Goldman (1884–1971) in 1926 [Strikha, 2024; Kozyrski, 2020]. Isaak Khalatnikov, who was in Ukrainian school during that period remembers [Teryaeva, 2019]: "As for the general "Ukrainization" of those years, we probably didn't feel it at all. We had to learn Ukrainian, and we did. It wasn't seen as a disaster, unlike today's attitudes toward such reforms. Ukrainian is a very resonant language and easy to learn." Ukrainian in school education was still prominent after the second world war, as Kharkiv-born Mark Azbel (1932-2020) discovered on coming back to Kharkiv in 1944 "All courses were easy for me except for the Ukrainian language. … I was not the only failure; Kharkov having just been liberated, there were other children who were also ex-refugees and who had been in Siberia, or other regions outside the Ukraine, where this language was not studied in school." [Azbel, 1981]. In 1949, Bogolyubov (1909-1992) published his Lectures on Quantum Statistics in Ukrainian at Kyiv university publishers [Gerasimenko, 2026].

[22] Since 1892. Collection of the Mathematical, Natural History and Medical Section, where the main body of natural science publications has been published, has existed since 1897.

**Table 1** Major periods of Ukrainization and Russification based on [Dzyuba]. Not indicated are periods of partial "softening" of Russification, e.g. 1905-1910, Khrushchev Thaw (1953-1958) and perestroika (1985-1991).

| Ukrainization | Russification |
|---|---|
| | **Imperial Russification** (up to 1917)<br>Restrictions[23] on Ukrainian language and culture (e.g. total ban for usage of Ukrainian in administration, education, severe restrictions on Ukrainian fiction and theatre, etc.); promotion of Russian in all spheres of public life; perception of Ukrainian identity as a threat. |
| **Revolutionary Ukrainisation** (1917–1921)<br>Short revival during a brief period of Ukrainian independence; expansion of Ukrainian language in administration and education. | |
| **Early Soviet Korenizatsiya** (1923 – 1931)<br>Official policy of Ukrainization promoting Ukrainian language and culture in education, administration, and media as the instrument of the "building of socialism"; widespread Ukrainisation of schools and publications. | |
| | **Stalinist repression** (1932–1953)<br>Abrupt end of Ukrainisation policy, purges of Ukrainian elites, Holodomor (1932-33), suppression of language reforms. Russian language strengthened as the main language of administration, education, and upward mobility. |
| | **Late Soviet Russification** (1958 – 1985)<br>Education reforms favoring Russian; decline in Ukrainian-language publications; Russian becomes dominant in cities and higher education. |
| **Independence** (from 1991)<br>Restoration and promotion of Ukrainian as the state language; gradual expansion of Ukrainian in public life, education, and media (varies over time). | **Partial Occupation by Russia** (from 2014)<br>Russia's policies on the temporarily occupied territories of Ukraine are similar to those described in "**Stalinist repression**". All the forms of Ukrainian-language public activities there are forbidden and persecuted. |

[23] The Valuev Circular of 1863 and the Ems Decree of 1876.

This progress was quite short-lived, although speaking Ukrainian was never banned and publishing in Ukrainian has not been banned since 1917, it faced systematic suppression through extensive periods of Russification (prior to 1917 and between 1932-1985) shown in **Table 1**. It was often implied that in the Ukrainian "language there was neither science nor scientific terminology" (although, in fact, it was not so at least since the 1920s) compared to "Russian, which had a large scientific literature" [Timoshenko, 1968, p. 290]. The russification policy, combined with regular purges of Ukrainian elites[24] in the Soviet Union (1930s, 1960s) led to dominance of Russian ideology and language not only in the Ukrainian schools but also in its scientific institutions.

"In the cities of Ukraine in 1958, only 21% of pupils studied in Ukrainian schools (in 1926 - 97%)" [Dzyuba]. Another example is provided in [Strikha, 2024]: the Ukrainian Journal of Physics was published from 1956 only in Ukrainian (with extended English summaries). Then, in 1965, "for the convenience of readers from other republics" a parallel Russian version was launched. At the end of 1976, the Ukrainian version was eliminated "as unnecessary". In the 1980s, Ukrainian-language physics seemed to come to an end, but the situation was reversed by the national revival during perestroika and independence, which came after it: by the mid-1990s the journal again became Ukrainian-language, with English edition appearing in 2004 [Strikha, 2024].

The 2001 census of the population of Ukraine[25] [SSCU, 2001] recorded that 67.5% spoke Ukrainian and 29.6% Russian. Those citizens of Ukraine who considered themselves Ukrainian spoke 85% Ukrainian and 14.8% Russian, while those who considered themselves Russian spoke 3.9% Ukrainian and 95.9% Russian. These figures had changed gradually during the last decades, especially since the beginning of a full-scale Russian-Ukrainian War in 2022. All the polls of the last years demonstrate more than 90% of Ukrainian speakers on the government controlled territory. These changes are caused both by the real switch in the language practices and by the fact that appr. 20% of national territory, where Russian had dominated in 2001, are no longer under Ukrainian control.

The use of Ukrainian language in scientific publications has long served as a marker of national identity, despite systemic restrictions in the Russian Empire and the USSR. From the first works of Ivan Pulyui in the journals of the Shevchenko Scientific Society to the period of "korenizatsiya" of the 1920s, there was an active formation of terminology and a publishing base. However, waves of repression against the elites and the introduction of Russian as the language of "higher science" led to the gradual displacement of Ukrainian from academic circulation. Today, despite the lasting consequences of Russification, independence and national revival have returned the Ukrainian language to the status of the state language in science, which is critically important for the international identification of the Ukrainian scientific community. However, the effects of the decades of these language policies are felt among scientists from Ukraine even today, 35 years after the declaration of Ukrainian independence. The extensive use of the Russian language by scientists from Ukraine and Ukrainian scientific diaspora still obscures their identification as Ukrainians abroad.

[24] Notably Lev Strum (1890-1936) [Kragh, 2025], Lev Shubnikov (1901-1937) and Matvei Bronstein (1906-1938) were executed. Dmytro Ivanenko (1904-1994), Oleksandr Leipunsky (1903-1972), Oleksandr Goldman (1884–1971), Vadym Lashkaryov (1903 - 1974) and many others were imprisoned.
[25] The 2001 census was the most recent, and the only census conducted in independent Ukraine to date.

### 5. Impact of Russification on population of Ukraine

In addition to actions to promote the use of the Russian language detailed above, the Russification policy included various other campaigns of integration of Ukrainian population:

- through encouragement of cross-marriages [Anna Ioannovna, 1905], typical outcomes of which were aptly documented in the autobiography of Gamow, born in Odesa: “one of my paternal forebears was an officer in the Imperial Russian Army and was sent from St. Petersburg (now Leningrad) some time early in the eighteenth century to liquidate the unruly Zaporozhets[26]… On my mother’s side there is a long line of southern clergy who claimed descent from the Zaporozhets… ” [Gamow, 1970];
- through encouragement and facilitation of movement of population from other regions of the Russian Empire to Ukraine [Plokhy, 2015], which intensified during the Soviet Union following depopulation after Holodomor, WWII, etc.: for example, the family of Bogolyubov moved to Kyiv in 1913, when his father was hired to teach in Kyiv University [Bobolyubov, 1996];
- through imperial [Alexander III, 1899] and Soviet resettlement programs, large numbers of Ukrainians migrated both to imperial centers and to distant regions of the Russian Empire and the Soviet Union, including the Far East. In the 1959 census of the USSR, the population of Ukrainians in Russia[27] was about 3,359k people [USSR, 1959], reducing slightly to 3,346k by 1970 [USSR, 1970], driven by Russification, assimilation[28], displacement and emigration, discussed in the next section. For example[29]:
    - P. Kapitza (1894-1984) was born in Kronstadt near St. Petersburg; his mother came from the noble Stebnytsky family of the Volyn region in Ukraine, while his father was from Bessarabia.
    - I. Tamm (1895–1971) was born in Vladivostok, although his family was from Yelisavetgrad (now Kropyvnitsky) and he finished school and gymnasium in Yelisavetgrad, then studied in Edinburgh and Moscow; from 1918, he taught at Tavria University in Simferopol, and in 1921–1922 at Odesa Polytechnic; from 1922, he worked in Moscow.

---

[26] Cossacks of the Zaporozhian Host established a semi-autonomous political entity in southern and central Ukraine, that challenged the authority of the Polish–Lithuanian Commonwealth, the Tsardom of Russia, and the Crimean Khanate (16th–18th century). As a military and political force that resisted foreign domination they were inspirational for the establishment of Ukrainian statehood and national identity. [Plokhy, 2015]

[27] Henceforth used to denote the Russian Federal Socialist Republic (1921-1991).

[28] Children of mixed marriages in the USSR were allowed at the age of 16 to select their ethnicity and many preferred to declare themselves as Russian on their internal passports. They had no choice if both parents were the same ethnicity. This information was also recorded in the USSR on the birth and marriage certificates. This information is no longer recorded in the Ukrainian passports and birth certificates.

[29] More examples were detailed footnotes 15-17.

- V. Veksler (1907-1966) was born in Zhytomir, his family moved to Moscow in 1915. Abraham Zelmanov (1913-1987) was born in Hadyach (Poltava region), his family moved to Leningrad in 1924.
- I. Shklovsky (1916-1986) born in Hlukhiv, Sumy region, his family moved to Kazakhstan in 1930.
- G. Budker (1918-1977) was born in the Ukrainian People's Republic in 1918, after education and work in the Moscow region (1941-1958) went on to establish the Institute of Nuclear Physics in Novosibirsk, which now carries his name [Johnsen, 1977].

The results of Soviet policy of Russification can be clearly seen in the population census numbers in **Table 2**: Russian population in Ukraine grew steadily and doubled between 1928 and 1989, while the percentage of Ukrainian population in Ukraine decreased in the same period. The increase in the percentage of Ukrainians in 2001 is not related to the increase in the actual number of people, but rather an artifact of the decrease of the total population. At the same time Jewish population of Ukraine, which was the third biggest in the world in the late 1920s after that of the USA and Poland [Linfield, 1928], collapsed due to Holodomor, Holocaust, Soviet assimilation policies and population displacement within the USSR as well as emigration waves discussed in the next section.

It is also important to note the urbanisation of the population of Ukraine after 1928. Despite the fact that in 1900 more than 90% of ethnic Ukrainians were living in rural areas, and the cities on Ukrainian territory had (for administrative reasons of the imperial policy) mostly the mixed Russian-Jewish-Ukrainian-Polish population, in the 1950s both the majority of Ukrainians lived in urban areas, and the majority of urban areas population in the Ukrainian SSR were now ethnic Ukrainians.

**Table 2.** Percentage of Ukrainians, Russians and Jews in the population of Ukraine based on census results for 1928 [Klyovanyy, 1928] (only within the borders of Ukrainian SSR at the time), 1959 [USSR, 1959], 1970 [USSR, 1970] and 1989/2001 [SSCU, 2001].

| **Total population** | **28,900,000** | **41,869,046** | **47,126,517** | **51,917,817** | **48,457,000** |
|---|---|---|---|---|---|
| **% of total (urban only) population** | **1928 (urban only)** | **1959** | **1970** | **1989** | **2001** |
| **Ukrainians** | 79% (38.5%) | 76.8% | 74.9% | 72.7% | 77.8% |
| **Russians** | 9.5% (27%) | 16.9% | 19.4% | 22.1% | 17.3% |
| **Jews** | 6.5% (29.5%) | 2.0% | 1.7% | 0.9% | 0.2% |
| **Urban** | 18.2% | n/a | n/a | 65% | 63% |
| **Rural** | 81.8% | n/a | n/a | 35% | 37% |

The policy of Russification caused profound demographic transformations in Ukraine by encouraging interethnic marriages, mass resettlement, and assimilation. Statistical data indicate a rapid increase in the share of the Russian population (from 9.5% in 1928 to 22.1% in 1989) against the backdrop of the migration of Ukrainian intellectual elites to imperial centers. The examples of Kapitsa, Tamm, Veksler and Budker illustrate the scale of migration of these scientists, whose Ukrainian roots were often obscured by the Soviet system. At the same time, there was a profound shift in the national composition of the urban population and the tragic disappearance of the Jewish community, which fundamentally changed the socio-ethnic landscape of the country and the conditions for the development of Ukrainian science.

## 6. Emigration from the territory of Ukraine

In addition to the population displacements within the territory of the Russian Empire/Soviet Union prior to 1991, which were discussed in the previous section, in the past 150 years there were six main waves of emigration from the territory of Ukraine summarized in **Table 3**.

The first wave of emigration (1880-1915) was driven by the agricultural workers, artisans and small traders escaping from economic hardship, rapid industrialization and pogroms [Rovenchak, 2018; Petrovsky-Shtern, 2014, p. 18]. It is not surprising that there are no names of established scientists during this wave. The biographies of M. Ostrohradskyi (1801-1862), I. Pulyui (1845-1918) and S. Timoshenko (1878-1972) illustrate how easily scientists could travel across Europe for study, research, or leisure during that period, benefiting from the integrated global economy. Emigre scientists and engineers from this wave typically left Ukraine as small children or were born abroad and did most of their studies abroad. The end of this wave was driven by departures to escape the effects of WWI with its military draft.

The second emigration wave from Ukraine was driven by people trying to escape the Soviet rule after the collapse of the Russian Empire in 1917, the Austro-Hungarian Empire in 1918, as well as by the failure to establish Ukraine as an independent state. S. Timoshenko [Timoshenko, 1968], E. Ginzton [Cottrell, 1995], and G. Charpak [Charpak, 2008] provide vivid accounts of this period, describing the difficult decisions they and their families faced as they left their homes—sometimes not once, but twice—and the complex journeys they undertook to reach their destinations. Some emigres considered themselves Russian and preferred to speak Russian (even if they knew Ukrainian), those who were more “nationally conscious” [Rovenchak, 2018], not only considered themselves Ukrainian and preferred to speak in Ukrainian, but also created extensive Ukrainian diaspora organization[30], especially strong at that time in Czechoslovakia, where it was supported by president T. Masarik.

[30] Examples of these efforts are Free Ukrainian University (from 1921) [Petrovsky-Shtern, 2014, pp. 297-311], the Ukrainian Economic Academy in Poděbrady (1922-1935) [Zavorotna, 2020], etc. These organizations aimed to educate professionals for work in independent Ukraine once the Soviet rule collapsed.

**Table 3.** Main waves of migration from the territory of Ukraine based on [Rovenchak, 2018; Petrovsky-Shtern, 2014]. “*” means departure from Moscow/St. Petersburg. ^ through family. If date is not given, the family moved shortly before birth.

| Timespan and main destinations | Reasons | Representative scientists (year of departure abroad) |
|---|---|---|
| 1st wave (1880-1915) to USA, Canada, Brazil, Argentina, Palestine, South Africa | Economic hardship, pogroms, land shortage, political oppression under Austro-Hungarian & Russian Empires, war | H. Brown^, M. Chodorow^, S. Drell^[31], M. Goldhaber^ (1914) and G. Goldhaber^[32], G. Breit^ (1915)[33], L. Lederman^, M. Gell-Mann^[34] |
| 2nd wave (1919-1939) to Western Europe (Czechoslovakia, France, Germany, Poland), USA, Palestine | Wars, famine (Holodomor 1932–33), political persecution | G. Kistiakowsky (1920), S. Timoshenko (1919, 1920), G. Wataghin^ (1920), M. Haissinsky (1921)[35], M. Moshinsky^ (1924), E. Ginzton^[36] (1919*, 1927), A. Smakula (1922, 1930), A. Chichibabin (1930*), G. Charpak^ (1926, 1931)[37], G. Gamow (1933*), S. Deser^ (1935)[38] |
| 3d wave (1940-1955) to Germany, USA, Canada, Australia | WWII displacement, Nazi occupation, Soviet repression, forced labor | Z. Chraplyvy (1944), L. Romankiw^ (1944)[39], R. Hoffman^ (1949) |
| 4th wave (1970-1989) to the USA, Israel, etc. | Failure of dissident movement of 1960s | A. Vilenkin (1976)[40], M. Azbel (1977*)[41], E. Gliner^ (1980*) |
| 5th wave (1991-2021) to North America, Western Europe, Israel | Economic and professional opportunities, post-Soviet labor migration. Some of these scientists emigrated abroad through Russia, but only a few remained[42] in Russia after 2022. | E. Shuryak (1990*), G. Gorelik (1992*), V. Yarba (1992*), T. Hryn’ova (1997), V. Akulov[43] (1998), D. Kovalsky (1998*), A. Neronov (1999*), A. Boyarsky (1999*), M. Titov (2003*), M. Vyazovska (2005), S. Barsuk (2006*), L. Shchutska (2008*), O. Shadura (2014), V. Zhovkovska (2019)... |
| 6th wave (after 02.2022) | War | Ongoing |

[31] http://biographicalmemoirs.org/pdfs/drell-sidney.pdf
[32] Crease, 2012
[33] Hull, 1998.
[34] https://www.webofstories.com/play/murray.gell-mann/1
[35] Cottin, 1979.
[36] Cottrell, 1995.
[37] Charpak, 2008.
[38] Chang, 2023.
[39] Lazur, 2021.
[40] Stober, 2009.
[41] Azbel, 1981.
[42] For example, I. Yeletskikh has been based in Dubna since 2007.
[43] https://esu.com.ua/article-43550

The third wave was driven by refugee families from the western regions of Ukraine annexed by the USSR just before WWII and prisoners of the postwar refugee camps in Austria and Germany [Rovenchak, 2018; Kobchenko, 2020]. L. Romankiw remembers “I left Ukraine when I was 13 years old. When Soviet troops arrived in Zhovkva in the evening, lists of those who were to be shot the next morning appeared that night. My father, who worked as a lawyer in Zhovkva, was on those lists. He managed to be taken across the Solokia River to Ugniv, where the Germans happened to be detained, and that's how my father escaped to the German unit. And we - my sister, my mother, and I - were taken out the next night. For those who wanted to escape the Soviet occupation, a special unlit train ran to Ugniv at night. If I had stayed in Ukraine, I would have been in Siberia, or shot.” [Lazur, 2021]. The third wave, dominated by professionals (well-educated doctors, lawyers, engineers etc.) reinforced the efforts of the second wave of emigrants to preserve the Ukrainian national identity.

Following the failure of the dissident movement of the 1960s, the Six Day War (1967) and the invasion into Czechoslovakia in 1968, many activists from Ukraine realized that the Soviet system could not be reformed from the inside. As part of detente with the West, the Soviet government allowed the emigration mostly for the reason of “family re-unification” [Petrovsky-Shtern, 2014, p. 58-60; Azbel, 1981]. The members tended to be anti-Soviet. Due to the displacement of scientists born in Ukraine to the Soviet scientific centers in Russia, some of them were emigrating from there, as can be seen from Table 3. One of the outcomes of this emigration wave, which continued after 1991, was creation of a large Russian-speaking Jewish diaspora spread all over the world. Majority of them had family roots in Ukraine[44].

The following wave was driven by voluntary economic migration after the dissolution of the Soviet Union and [Rovenchak, 2018] notes its “specific language feature” : “a great number of Russian-speaking Ukrainians counting about a half of its representatives”. Many members of this migration wave relied on existing international connections and collaborations, which were biased towards Soviet centers of power. S. Zhuk estimated that “Between 1991 and 2005, the overwhelming majority (85 percent) of post-Soviet academic migration to the United States, at least, was Russian by origin…. More than 80 percent of that diaspora in the West consists of Muscovites, the residents of Moscow!”[45] [Zhuk, 2024]. Indeed, a non-negligible number of scientists born in Ukraine who settled abroad passed through institutions in Russia, either as students or researchers. A fraction of them “identify themselves not as Ukrainians (nationally, ethnically, culturally, by language), but rather as Russians, Russian-speaking, (post-)USSR-residents” [Rovenchak, 2018]. This is particularly the case for older generation scientists born in the Ukrainian SSR who settled Russia prior to 1991 and never lived in independent Ukraine. Russia’s invasion of Ukraine in 2014 cut off links between Russian and Ukrainian scientific institutions. In the meantime Ukrainian institutions developed direct foreign contacts bypassing former Soviet scientific centers (e.g. the Ukrainian associate status in the

[44] The census data confirms migration of Jews from Ukraine to the Soviet centers: while in 1928 1796k Jews resided in Ukraine and only 518k in Russia [Linfield, 1928], by 1959 Ukrainian Jewish population reduced to 840k, compared to 897k in Russia [USSR, 1959]. This trend was reinforced by larger numbers emigrating from Ukraine, than from Russia between 1970 and 1997 [Petrovsky-Shtern, 2014, p. 60].
[45] S. Zhuk’s study was focused on specialists in Russian studies (history, anthropology, political science) and might need to be updated for particle physics.

EU “Horizon-2020” program in 2015, Associate Membership of Ukraine in CERN in 2016, LIA/IRP IDEATE with CNRS, France [Grynyov, 2019], etc.).

Full-scale invasion of Ukraine by Russia on 24th of February 2022 initiated a new wave of emigration of scientists from Ukraine. According to very rough estimations more than 10% of active researchers had left Ukraine in 2022, part of them had returned later, when the situation became more predictable. This process is still ongoing and its impact is not clear at this point.

The history of emigration from Ukraine over the past 150 years has six main waves, each of which had a specific impact on the scientific diaspora. If the first wave was mainly economic, then the following ones (after 1917, 1945 and 1968) pushed out the intellectual elite due to wars and political repressions. Emigration enabled many scientists born in Ukraine, such as G. Breit, G. Charpak, G. Gamow, E. Ginzton, M. Goldhaber, G. Wataghin, to name a few - to make seminal contributions to the field of particle physics. An important problem remains the blurring of identity: due to the transit of many scientists through Russian centers before leaving for the West, they were for a long time perceived as part of the Russian-speaking or Soviet diaspora, which is only now starting to change thanks to Ukraine’s direct contacts with international institutions.

## 7. Criteria to define a scientist as Ukrainian one

The previous sections outlined the main difficulties in defining who should be considered a Ukrainian scientist based on the usual criteria. Nationality is problematic because Ukraine became independent only in 1991 (although had a formally sovereign status within the USSR since 1922 [Constitution UkrSSR, 1929, Art. I.3]) and experienced large population movements over the past 150 years. Ethnicity, culture, and language are also challenging criteria due to the country’s long-term policy of Russification.

Thus our study will apply at least one of the four basic criteria to define a scientist as Ukrainian:

1. **Origin:** Born within the present internationally recognised state borders of Ukraine[46] [VRU 1992].
2. **Affiliation:** Conducted research at scientific institutions in Ukraine.
3. **Scientific contribution:** Founded or developed scientific schools in Ukraine.
4. **Self-identification:** encompasses the Ukrainian cultural and political tradition, use of Ukrainian language, participation in Ukrainian scientific and cultural organizations[47], etc.

Prior to 1991, it was challenging and unrewarding for scientists to declare themselves Ukrainian, especially given the pressures of totalitarian systems. Furthermore, regular smear campaigns have been carried out against the Ukrainian independence movement since the 1920s, both abroad and within Ukraine [Petrovsky-Shtern, 2014, pp. 3–64]. Echoes of these campaigns can be found in the memoirs of well-known physicists [Weisskopf, 1991, p. 13; Charpak, 2008, pp. 9–11]. "Overall, through the entire Cold War, and even today, Soviet/Russian intelligence has attempted to persuade Western audiences that the Ukrainian diaspora and the entire Ukrainian movement for independence were organized by the followers of Stepan Bandera and by the "fascist" Ukrainian collaborators with Nazis during World War II, who fled to America and who were used by the American intelligence agencies in various American "imperialist wars" against the Soviet Union, and now against the Russian Federation." [Zhuk, 2024]. It is therefore particularly significant that both M. Planck and A. Einstein, leading figures in the international scientific community, chose to accept membership in the Shevchenko Scientific Society in 1924 and 1929, respectively [Holovatch, 2018]. Planck wrote in his letter of acceptance "I believe that you know that in Germany Ukrainian culture sparks lively interest and the Ukrainian political struggles are fully and loyally supported …" [Holovatch, 2018].

[46] Ukraine from here on assumes within the state borders of Ukraine as defined in [VRU 1992], even if the occurrence predates the quoted reference.
[47] Shevchenko Scientific Society, Ukrainian Academy of Sciences, Free Ukrainian University, Ukrainian Economic Academy in Poděbrady, etc.

Therefore one should be accurate in defining this or that researcher as a Ukrainian one, especially for the period before 1991, taking into consideration all the list of criteria presented above. E.g. Marian Smoluchowski can not be treated as a Ukrainian physicist, despite the fact, that all his principal results were obtained in 1898 - 1913 during the period of his work at Lviv university (then Austro-Hungary), because he had demonstrated all the time his strong Polish self-identification. Similarly Mykola Bogolyubov should not be regarded as merely a Russian scientist, despite his birth and death in Russia and his work in Dubna. His strong Ukrainian self-identification (including the fact of printing some principal results in Ukrainian in the time, when this was a kind of political demonstration), his long period of professional activity in Ukraine, and his strong involvement into Ukrainian R&D system, even when his main research position was in Russia, should be taken into account. The case of Lev Landau is more complicated: ethnic Jew, born in Baku (now Azerbaijan), he had obtained part of his principal results in Ukraine and part in Russia. However, his personal self identification was not Russian and not Ukrainian (although he knew Ukrainian while working in Kharkiv), but "Soviet". Therefore the definition "Soviet physicist", proposed by "Wikipedia" [https://en.wikipedia.org/wiki/Lev_Landau] seems to be the most proper one.

Defining a scientist as "Ukrainian" in a historical context requires a comprehensive approach based on four criteria: origin, affiliation, contribution to the creation of local scientific schools, and self-identification. This approach helps avoid simplistic interpretations caused by the long absence of independent statehood and the policy of Russification. The examples of Smoluchowski[48] (Polish identity), Bogolyubov (Ukrainian self-consciousness), and Landau (Soviet identity) demonstrate the importance of considering personal position and cultural context. Such a balanced approach helps restore historical justice to Ukraine's scientific heritage, countering long-standing efforts to diminish its national subjectivity in the eyes of the world community.

[48] B. in 1872 in Vorder-Brühl (near Vienna), worked in Lviv in 1899-1913, where he was president of the Polish Copernicus Society of Naturalists, (1906–1907). Moved to Krakow in 1913 and died there in 1917.

## 8. Particle physics pre-history

The first physics paper in Ukrainian in 1877 was about radiometers. Its author, Ivan Pulyui (1845-1918), born in Hrymailiv in Ternopyl region, was developer of *cathode lamp*s (called *Pulyui lamps*) which enabled him to discover X-rays in 1895 or even earlier independently from Wilhelm Röntgen[49] and produce high-quality X-ray images. Pulyui's observations confirmed that X-rays penetrated matter, affected photographic plates and had possible biological effects [Pulyui, 1996]. He also hypothesised in a series of articles "Radiant electrode matter" (1880-1882) that cathode rays are streams of negatively charged electrode particles [Holovatch, 2018], anticipating by more than a decade a discovery of the electron by J.J. Thomson in 1897 [Thomson, 1897]. Pulyui's experience highlights the importance of promptly publishing and publicizing one's discoveries and thus establishing scientific priority.

The question of scientific priority was well understood by Marie Curie (née Skłodowska), who received two Nobel Prizes (1903, 1911) for her research of the phenomena of radioactivity. Moise Haïssinski[50] (1898–1972), born in Tarascha in the Kyiv region, joined Marie Curie's laboratory in Paris as personal assistant in 1930. M. Haïssinski's research subsequently laid the foundations of modern radiochemistry[51] [Haïssinski, 1957].

The priority was also on top of the minds of physicists from UIPT, when they decided to search for the disintegration of atoms of lithium by high velocity protons[52], following visits to Kharkiv by Gamow and Cockcroft[53] in August 1931 [Lucie Str, 1944]. Such an experiment had been in preparation at Cambridge since 1929. It was based on the concept of quantum tunneling proposed by Gamow to explain nuclear alpha decay [Gamow, 1928; Novosyadlyj, 2026]. The experiment in Kharkiv was completed in October 1932 [Sinelnikov, 1932], a few months behind that of Cockcroft and Walton[54]. Nevertheless, K. Synelnykov, A. Leipunsky, A. Walter and G. Latischev announced their achievement on the front page of the main Soviet daily, *Pravda* on 22 October 1932[55]. This result put UIPT (now known as KIPT) on the front line of nuclear physics research in the Soviet Union.

Furthermore, in early 1940, "Kharkiv researchers Friedrich Lange, Volodymyr Shpinel, and Viktor Maslov filed patents for a nuclear bomb and methods for producing uranium-235 - predating similar projects elsewhere in the world. At the time, however, these ideas were considered too revolutionary and did not receive the support and development they deserved in the Soviet Union. Only after the atomic bombings that ended World War II did it become clear how advanced the research conducted at UIPT truly was. Despite this, Kharkiv's role is often overlooked in textbooks and publications on Soviet nuclear history, overshadowed by that of Moscow" [Hryn'ova, 2026].

---

[49] Röntgen received the Nobel Prize for the discovery of X-rays in 1895.
[50] His son, Jacques Haïssinski's (1935-2024), was a well-known particle physicist [CERN, 2024].
[51] Moise Haïssinski studied the formation and properties of colloidal solutions of radioactive elements, used electrochemical and radiochemical methods to investigate compounds of protactinium, polonium, and curium, established the fundamental principles governing the radiolysis of water and aqueous base solutions and predicted of the electron configurations, oxidation states, ionic and atomic radii, ionization potentials, and key chemical behavior of heavy elements [Cottin, 1979].
[52] As radioactivity is the process by which unstable atomic nuclei spontaneously decay, this experiment would result in a controlled version of similar phenomena.
[53] Gamow, Synelnykov, and Lev Landau—who was also working in Kharkiv at the time—were acquainted with Cockcroft from their earlier stays in Cambridge, UK.
[54] This publication earned Cockcroft and Walton the 1951 Nobel Prize in Physics [Cockcroft, 1932].
[55] No. 293 (5458) [Ranyuk, 2006, pp. 100-103]

Such erasure is exemplified by the story of the discovery of type II superconductors—now essential for the magnets of the Large Hadron Collider (LHC) at CERN—by Lev Shubnikov, Vladimir Khotkevich, Georgy Shepelev, and Yuri Rjabinin at the UIPT in the 1930s. Their names were banned following the Soviet purges and the discovery forgotten [Shepelev, 2011]. These materials were rediscovered again only in 1961.

The understanding of atomic and nuclear structure, along with the experimental techniques developed to study them, provided the foundation for particle physics research. At the same time, physicists gained considerable prestige within their governments in the 1950s, as their expertise had proven vital for the development of powerful new weapons and strategic technologies. This influence enabled them to secure funding for new facilities and to participate in intergovernmental negotiations, ultimately fostering international collaboration on the peaceful applications of particle and nuclear science, as discussed in the last section.

Scientists from Ukraine have been at the forefront of research aimed at understanding the nature of subatomic particles, through the pioneering work of I. Pulyui on cathode rays and X-rays, G. Gamow's groundbreaking insight into quantum tunneling to explain experimental results of alpha decay, as well as early projects on particle accelerators and studies on the use of nuclear energy in Kharkiv. Despite repression and the suppression of achievements (in particular, the discovery of Type II superconductors by Shubnikov), this period fostered highly qualified scientific personnel and laid the foundation for future research in particle physics.

## 9. Contributions to particle physics theory and related fields

Progress in particle physics relies on the close interplay between experiment and theory, with experimental discoveries revealing unexpected new phenomena and theoretical frameworks explaining them and predicting new ones, guiding further investigation, as was the case for the experiments on the artificial disintegration of lithium by fast protons. Furthermore, work in theoretical physics, required little more than training, paper, and access to ideas, whereas experimental physics depended on costly equipment, materials, and stable infrastructure, which were often scarce or tightly controlled under constrained economic conditions, making theoretical research far more accessible for physicists in scientific institutions in Ukraine and the USSR in general.

The leading center for theoretical particle physics research in Ukraine has long been the Kharkiv Institute of Physics and Technology. A comprehensive review of the work conducted there can be found in [Khramov, 1991]. The first head of the theory group from 1929 till 1931 was D. Ivanenko (b. Poltava 1906-1996), best known for proposing the proton–neutron model of the atomic nucleus [Schenderovsky, 2026]. Following his departure, from 1932 till 1937 the group was led by Lev Landau (b. Baku, Azerbaijan 1908-1968), who developed the concept of the "theoretical minimum". This rigorous examination system for training theory students produced a generation of physicists who made groundbreaking contributions to quantum field theory, particle and nuclear physics, solid-state theory... The first ones to pass the theoretical minimum were O. Kampaneyets (b. Dnipro 1914-1974), E. Lifshitz (b. Kharkiv 1915-1986), O. Akhiezer (b. Cherykaw, Belarus 1911-2000), and I. Pomeranchuk (b. Warsaw, Poland 1913-1966). While at Kharkiv, Landau with Lifshitz, Pyatigorsky and Rozenkevych began writing a ten-volume *Course of Theoretical Physics,* which for many years served as a basic physics textbook. After Landau's flight to Moscow, the Kharkiv school continued to thrive under the leadership of O. Akhiezer in particle physics and I. Lifshitz in solid-state theory. Among their achievements are

- development by O. Akhiezer and M. Shulga (b. Kharkiv 1947-2024) of the quasi-classical theory of coherent radiation of channeled and over-barrier electrons and positrons in crystals which has applications in accelerator physics [Sytov, 2022],
- development by M. Rekalo (b. Donetsk region 1938-2004) of the relativistic theory of polarization effects in high-energy electron–deuteron scattering and his calculations with O. Akhiezer of electromagnetic characteristics of hadrons and [Neklyudov, 2004],
- as well as the contributions of D. Volkov, V. Akulov, and V. Soroka to the construction of theories of supergravity and supersymmetry [Nurmagambetov, 2026]. Theory groups of V. Ogievetsky (b. Dnipro 1928-1996) at JINR in Dubna and of A. Kapustnikov (b. Baku, Azerbaijan 1945-2003) and V. Skalozub (b. Dnipro 1949) at the University of Dnipro also worked in similar areas.

Another leading center of theoretical particle physics research in Ukraine was in Kyiv, centered around M.M. Bogolyubov [Gerasimenko, 2026; Gorenstein, 2026]. The Bogolyubov–Parasyuk theorem [Bogolyubov, 1955; Bogolyubov 1957], which was derived in 1955 at the Institute of Mathematics of the Academy of Sciences of the Ukrainian SSR with O. Parasyuk (b. Lviv region 1921-2007) [Zagorodnii, 2022], is one of the cornerstones of modern quantum field theory, providing a rigorous justification for the renormalization of divergences in Feynman diagrams. The establishment of the Institute of Theoretical Physics by Bogolyubov in 1966 [Gorenstein, 2026] attracted to Kyiv prominent physicists such as

- B. Struminsky, known for his work on the quark structure of hadrons and, in particular, his 1965 proposal of a new quantum number - color [Tkachov, 2009],
- as well as G. Zinovjev, who was a key figure for early integration of the Ukrainian particle physics community to CERN [Grynyov, 2026] and the development of relativistic heavy-ion physics in Ukraine [Gorenstein, 2026], in particular, studies of quark-gluon plasma[56].

Current research at the institute is focused on studies of new phase structures of matter at high energy density [Gorenstein, 2026] and low-x physics [Martynov, 2019].

[56] The term "quark-gluon plasma" was proposed in 1978 by E. Shuryak (b. Odesa 1948).

The prominent representatives of the Lviv school of theoretical physics are Zenon Chraplyvy[57] (b. Ternopyl region 1904-1972) and Vasyl Miliyanchuk (b. Dobrovidka 1905–1958). They contributed to early theoretical work in quantum mechanics and electrodynamics.

Many important theory contributions were also made by scientists working outside Ukraine, only a few are highlighted below. Gregory Breit (b. Mykolaiv 1899-1981) is known for the Breit–Wigner formula describing resonance behavior in nuclear and particle physics, his work on relativistic corrections in atomic structure. He also described a very rare process in which a positron–electron pair is created from the collision of two photons; this so-called Breit-Wheeler process was observed recently at the LHC [CMS, 2025]. Iosif Khriplovich (b. Kyiv 1937–2024) is known for his early work on Yang–Mills theory, discovering anti-screening in 1969 - a key ingredient of what later became known as asymptotic freedom, as well as studies of the weak interaction between electrons and nucleons. His theoretical predictions were confirmed by multiple experiments [Shifman, 2025].

Important theoretical research on the theory of relativity was performed at the Institute of Physics in Kyiv, by Lev Strum (b. Kyiv 1907-1936), who worked, in particular, on faster-than-light particles [Kragh, 2025], and Nathan Rozen (worked in Kyiv 1936-38) [Scherbak, 2012]. One should also note the foundational work of Matvei Bronstein on quantum gravity, of George Gamow on the Big Bang model and predicting primordial nucleosynthesis. Their work was extended further by Petro Fomin (1930-2011) [Bakai, 2020] who worked at KIPT and BITP. In 1973 he proposed that the Universe could spontaneously originate from a gravitational instability of the quantum vacuum, unifying general relativity with quantum field theory and laying an early foundation for quantum cosmology, a field further developed by Erast Gliner (b. Kyiv 1923–2021) and Alexander Vilenkin (b. Kharkiv 1949).

Under constrained economic conditions in Ukraine, theoretical research—requiring fewer resources—was often more accessible and therefore especially prominent. Major centers such as the Kharkiv Institute of Physics and Technology and scientific institutions in Kyiv, Lviv and Dnipro fostered influential schools of theoretical physics. Physicists, who were born or worked in Ukraine, played a significant role in advancing areas ranging from nuclear and particle physics to cosmology, shaping key concepts such as supersymmetry and supergravity, asymptotic freedom, and the Big Bang model.

## 10. Contributions to accelerator development

Experimental verification is a crucial part of establishing any theory. In particle physics, such tests often rely on particle accelerators. The field of accelerator physics developed rapidly after its emergence in the 1930s.

In 1944, V. Veksler[58], independently of Edwin McMillan, discovered the phase stability principle, which made modern synchrotrons and high-energy accelerators possible. He also played a leading role in the construction of the 10-GeV proton synchrotron at the Dubna laboratory, which now bears his name.

[57] He left Ukraine in 1944. Zenon's son, Andrew Chraplyvy (b. St. Louis, Missouri, USA 1950) is famous for his work on the development of high-capacity optical fiber communication systems.
[58] "Route V. Veksler" is the only CERN street named after a scientist born in Ukraine.

The development of high-power microwave devices like the cavity magnetron for radars during WWII [Cottrell, 1995; Yanovsky, 2016] made it possible to generate intense, short-wavelength electromagnetic fields, which initiated studies on their application to accelerate charged particles in linear accelerators (linacs). They were notably used in the linear electron accelerator Mark-I at Stanford University, which reached an energy of 6 MeV in 1947. Stanford linear accelerators are usually connected to the names of Bill Hansen, Sigurd and Russell Varian and W.K.H. Panofsky. What is overlooked is that the first director of the Stanford Linear Accelerator Center (SLAC) was Edward Ginzton[59] (1915-98), who was born in Dnipro. Ginzton and Marvin Chodorow[60] were instrumental for the technical breakthroughs which allowed to increase klystron power by over three orders of magnitude, which led to construction of linacs Mark II (49 MeV, 1951), Mark III (1 GeV in late 1950s), Mark IV (80MeV, 1955) and SLAC 2-mile linac (20 GeV in 1966, extended to 53 GeV in 1989). More details on these developments can be found in [Cottrell, 1995]. These machines led to an extensive particle physics program resulting in four Nobel Prizes: in 1961 for studies of electron scattering in atomic nuclei, in 1990 for confirmation of existence of quarks, in 1976 for charm quark discovery and in 2008 for precise measurement of CP violation in B mesons [Panofsky, 2007].

The program of the electron linear accelerator with beam energy of up to 2 GeV was initiated by A. Walter at KIPT in Ukraine in the 1950s [Ranyuk, 2006]. Its construction was overseen by Ya. Fainberg[61] who suggested and realized a fundamentally new method for providing the simultaneous radial and phase stability of particles in linear accelerators on the background of accelerating fields, the method of alternating-phase focusing [Fainberg, 2005]. KIPT 2 GeV linac operated in 1965 to 1990 and was the largest electron accelerator in Europe—and for a few months, the largest in the world—until the Stanford linear accelerator became operational [Ranyuk, 2006]. It should be noted that the Kharkiv program remained mostly invisible to scientists abroad, as “[m]ost of KIPT results up to 1990 were published in Soviet journals” [Levchuk, 2026]. Although M. Swartz remembers that “the great polarization problem in 1993” run of SLD in SLAC [Swartz, 2025] was resolved thanks to a pre-print from Kharkiv [Levchuk, 1992].

Panofsky noted on his first visit to the USSR in 1956 that “we saw no experimental setups at either ITEP or Dubna that were not patterned after prior Western work. An exception to this observation was Budker’s laboratory in Moscow, which was undertaking pioneering efforts in some daring new technologies: plasma confinement experiments, pulsed magnets with boundaries defined by eddy-current sheets, and other novel technologies.” Panofsky's visit led to long-term collaboration between the institute which Budker founded in Novosibirsk in 1959

[59] Ginzton was also CEO of Varian Associates (1964-72), one of the founding companies of the Silicon Valley in California. The Microwave Laboratory at Stanford, where he worked now carries his name.
[60] M. Chodorow’s family came to the USA from Ukraine shortly before his birth [Chorodow, 2005]. Chorodow was also instrumental in the creation of the Applied Physics Department at Stanford University.
[61] Born in Zolotonosha, Cherkassy region in 1918. Died in 2005. [Fainberg, 2025]

and SLAC [Panofsky, 2007]. Gersh Budker is most known for his invention of the electron cooling method which is widely used in modern accelerator facilities all over the world. In Novosibirsk, he initiated construction of multiple storage rings: VEP-1 (1963), electron-positron VEPP-2 of 2x700MeV (1967), VEPP-3 2x2.2GeV (1972) and VEPP-4 2x7GeV (1985). These machines hosted experiments on meson resonances [Johnsen, 1977]. Other scientists from Ukraine, such as V. Baier[62] (b. Kharkiv 1930-2010), worked with G. Budker on the development of electron-positron colliders in Novosibirsk.

To finance their research, accelerator scientists and engineers at Kharkiv, Stanford and Novosibirsk constructed multiple other smaller linacs for use in particle physics, medicine, and industry [Cottrell, 1995; Ranyuk, 2006; Johnsen, 1977]. Ukrainian factories provided components for big international accelerator projects, e.g. ZAPOROZHTRANSFORMATOR from Zaporizhzhia delivered transformers for LEP200 at CERN in 1990-1992 [CERN, 1992].

In recent years the accelerator research in Ukraine was centered at the Institute of Applied Physics of the National Academy of Sciences of Ukraine in Sumy, which was founded and led by Volodymyr Storizhko (b. Olhovatka, Kharkiv region, 1935). Physicists born in Ukraine, such as Viktor Yarba [Yarba, 2019] (b. Dolyno-Kamyanka, Odesa region, 1935), as well as younger generation, Iaroslava Profatilova, Andriy Nosych, Yevgeniy Ivanisenko, Anna Vnuchenko, Volodymyr Rodin, continue to contribute to the development of accelerators all over the world.

The development of accelerator physics was the result of a synergy of fundamental ideas and engineering breakthroughs, where scientists of Ukrainian origin played a significant role. The discovery of the principle of phase stability by Veksler, the development of powerful klystrons by Ginzton at Stanford, and the creation of unique methods of focusing and cooling particles by Fainberg and Budker laid the foundation for modern colliders. The activities at KIPT in Kharkiv, the Budker institute in Novosibirsk and SLAC in the USA demonstrate the parallel progress in the development of electron accelerators, which led to a number of Nobel Prizes. Today, the integration of Ukrainian specialists and industrial capacities into global projects (in particular at CERN) ensures their continuing contribution to cutting-edge accelerator physics worldwide.

## 11. Contributions to experimental particle physics and related fields

The main experimental research program in nuclear and particle physics in Ukraine was initiated by Anton Walter (1905-1965) and centered on the KIPT 2 GeV electron linac, which was in operation in 1965-1990. Pavel V. Sorokin [ESU Sorokin] led the program, which was focused on studies of electromagnetic processes with nucleons and nuclei [Khramov, 1991; Ranyuk, 2006]: e.g. “results obtained by Kharkiv physicists on the single-pion photoproduction on the proton have been included in the Particle Data Group (PDG) compilations” [Levchuk, 2026].

[62] Left Ukraine to do PhD with I. Tamm (defended in 1958). His paternal half-sister, Nina Baier, is mother of A. K. Geim (the Nobel Prize in Physics 2010). His father Mykola Baier was the founder and first rector of Kamianets-Podilsky University, established as a stronghold of Ukrainian-language education within a short period of Ukrainian independence in 1918 [ESU Baier].

But the influence of scientists from Ukraine was global. Textbooks of Landau, written for and with his Ukrainian students, trained generations of physicists all over the world. The work of Gleb Wataghin[63] in Brazil in 1934-48 was transformative to the establishment of experimental and theoretical particle physics programs there [Dobrigkeit, 2026], in particular an extensive research program on comic rays. Laszlo Jenkovszky (b. Uzhgorod 1942 - d. Kyiv 2025) is credited with founding a school on forward physics in Hungary [Csörgő, 2025].

One of the most-known experimental particle physicists born in Ukraine is Maurice Goldhaber (b. Lviv, 1911-2011). In 1934, while doing PhD at Cambridge University, he performed his first notable experiment, the photodisintegration of the deuteron, which allowed a measurement of the mass of the neutron. With his wife Gertrude Scharff-Goldhaber, he established in the 1940s that beta particles are identical to electrons. His other major discovery was the determination of the helicity of the neutrino, in 1958, which contributed a key piece to the emerging theory of the weak interaction. M. Goldhaber was director of Brookhaven National Laboratory in the US from 1961 to 1973 and a patriarch of a family of three generations of US physicists [Crease, 2012].

Although there have been no particle physics experiments in Ukraine since 1991, Ukrainian scientists participate in a wide range of international particle physics experiments: at CERN [Gorenstein, 2026; Levchuk, 2026], at SLAC, at KEK, at DESY, at Fermilab,… They also made important contributions in the detector development, starting from Georges Charpak[64], who while working at CERN in the late 1960s, revolutionized modern particle physics detectors by inventing the multiwire proportional chamber, enabling fast, precise, and electronic tracking of particles in high-energy experiments.

Advanced ultralight interconnection technologies from RPE LTU in Kharkiv are used in the ALICE detector at CERN [Gorenstein, 2026]. In 1992, Ludmyla Nagorna from the Institute for Single Crystals in Kharkiv, proposed to use lead tungstate crystals ($PbWO_4$) for the ALICE and CMS detectors at the LHC, CERN [Tupitsyna, 2026]. Calorimeter measurements were crucial to the discovery of the Higgs boson by the CMS experiment, which was recognized by the 2013 Nobel Prize in Physics.

Kharkiv scintillators are also collecting data for the electromagnetic calorimeter of the Belle II detector at KEK (Japan), FermiLAT space telescope, and hadronic calorimeter of the CMS experiment at CERN. They were used in the electromagnetic calorimeter of the BaBar experiment at SLAC (USA) and in the Minimum Bias Trigger System of the ATLAS detector at CERN. The precise measurement of CP violation in B mesons by the BaBar and Belle experiments resulted in the 2008 Nobel Prize in Physics [Tarasov, 2026].

Recently, Kharkiv Institute for Scintillation Materials (ISMA) in collaboration with KIPT and Kyiv University (TSNUK) have contributed to technology proposal, simulation, prototyping, beam tests and innovative aging studies for the new luminometer (PLUME) for the LHCb detector at CERN. ISMA also coordinated Horizon Europe TWISMA project (2022-2025) with the participation of CERN and Institute of Light and Matter, University Claude Bernard Lyon, France, focused on research on innovative calorimeters for future high energy physics based upon advanced scintillation materials [Sidletskiy, 2025].

[63] Born in Birzula, now Podilsk, Odesa Region in 1899, died in 1987 in Turin, Italy.
[64] "Né en 1924, dans une bourgade de la plaine ukrainienne/Born in 1924, in a small town on the Ukrainian plain" writes Charpak in his autobiography [Charpak, 2008, p. 9]. He received the Nobel prize for his work in 1992.

Particle physics experiments also require storage and processing of huge quantities of data. The total LHC data over the last decade exceeds hundreds of petabytes, distributed across the Worldwide LHC Computing Grid (WLCG) for analysis by thousands of physicists globally.

Lubomyr Romankiw[65], co-inventor of the modern hard-disk magnetic storage technology based on electroplated thin magnetic films (developed with IBM), enabled high-density data storage that underpins modern computing. His innovations have been essential for storing, managing, and analyzing the massive datasets produced by contemporary particle physics experiments.

Scientists born in Ukraine, like Nikolay Govorun[66], actively worked on the automation of data processing for accelerator experiments in CERN and JINR. His legacy is honored at JINR with the "Govorun" supercomputer, a major scientific computing platform used for simulations and data analysis in experiments in Dubna.

KIPT in Kharkiv hosts Ukraine's Tier-2 computing center[67] within the Worldwide LHC Computing Grid, which pulls globally distributed computing resources to process and analyze the enormous volumes of data produced by LHC experiments [Levchuk, 2026].

Experimental particle physics owes fundamental discoveries and technological breakthroughs to scientists of Ukrainian origin. Maurice Goldhaber's contribution to the determination of the neutron mass and neutrino helicity, as well as the invention of the multiwire proportional chamber by Georges Charpak, became milestones in world science. The discovery of the Higgs boson at the LHC in CERN and the CP violation in B-mesons at flavor factories in KEK, Japan and SLAC, USA, were made possible thanks to Kharkiv scintillators. Lubomyr Romankiw's developments in the field of magnetic data recording and the participation of Ukrainian computing centers in the WLCG network provide processing of colossal volumes of information necessary to improve our understanding of the world around us.

## 12. Scientific diplomacy, instead of conclusion

Veksler is reported to have said "There are now three branches of physics: experimental physics, theoretical physics, and diplomatic physics" [Panofsky, 2007]. Indeed, Veksler, Budker, Bogolyubov, Joffe and other scientists from Ukraine were crucial for the opening of Soviet science to the world from the first Atoms for Peace Conference held in Geneva in 1955 [Lock, 1975; Lalli, 2024]. In 1956, at the Geneva Conference on High-Energy Physics, Ya. Fainberg, presented his ideas of a fundamentally new direction in the physics of accelerators—methods of collective acceleration. His colleague from KIPT, D. Volkov, was able to also visit CERN multiple times in the following years [Nurmagambetov, 2026] as well as the younger generation of Ukrainian physicists, who were sent to CERN by JINR for long-term stays: V. Yarba, N. Govorun [Yarba, 2019]… A systematic investigation of institutional archives and memoirs is needed to evaluate the permeability of the Iron Curtain for particle physicists from Ukraine, in particular to compare the level of cooperation to that of first globalization (1870-1914), which created the field. It is also worth recalling that Gregory Breit[68], a scientist from Ukraine who came to the United States as a teenager, proposed in April 1940 to the U.S. National Research Council that American scientists censor their work to prevent its potential military use by enemy powers during WWII—an idea that echoes modern calls for scientific sanctions.

---

[65] 1931-2024. Born in Zhovkva, Lviv region.
[66] 1930-1989. Born in Luhansk region in 1930, in Moscow University undergraduate student 1948-53 and PhD 1955-1961, worked in Kharkiv 1954-55 and at JINR from 1958.
[67] Another Tier-2 computing center was based at BITP in Kyiv, to process data for the ALICE experiment. It was managed by Evgenij Martynov (1948-2025).

In addition to scientific cooperation on accelerator and detector projects, Ukrainians had to face each other across the table during the Cold War negotiations on Arms control, with George Kistiakowsky[69] as member and chair of the USA President's Science Advisory Committee[70] in 1957-1961. Did they remember their common roots and the common experiences of the war atrocities which separated them at those meetings? Or were their Ukrainian identities completely overwritten or erased?

More expert studies are required to understand the impact on the community of particle physics from Ukraine of the declaration of independence in 1991, when for those in Ukraine travel across the Iron Curtain became “free” but the privileged status and most of the funding sources disappeared. Was this offset by the freedom of speech and ability to present themselves as Ukrainian scientists? Newly accessible CERN archive documents provide a detailed record of steps taken by the government of independent Ukraine to maintain the participation of its scientists at CERN [CERN, 1992; Grynyov, 2026].

We hope that this work will open a new chapter in the understanding of how scientists from Ukraine shaped the field of particle physics, which will help us not only to recover forgotten names but also to record the new ones, providing inspiration for future generations of particle physicists.

**Acknowledgements**

We are grateful to all attendees of the Workshop on “Ukrainian contributions to particle physics: history and prospects” for their active participation and lively discussions which were inspirational for this work. We also thank those who answered our multiple questions, send their comments, corrections and suggestions. One of us (T.H.) is grateful to the staff of CERN library for their help in having access to a large body of literature listed in the references as well as their patience through multiple renewals of these books. Some of them were purchased specially following our quest in the context of this study and we hope this work will raise interest in them. We are also grateful to colleagues at ISMA for their help in the publication of this work. All the mistakes which remain are ours.

[68] G. Breit led the early US work on the atomic bomb.
[69] 1900-1982. Born in Kyiv, became a US citizen in 1933, worked on the Manhattan project.
[70] This committee was created on November 21, 1957, by the US President Eisenhower in response to the launching of the Sputnik satellites. The Soviet Sputnik program was coordinated by S. Korolyov (1907-1966), born in Zhytomyr in Ukraine.

## REFERENCES

Alexander III. (1889). On the Voluntary Resettlement of Rural Inhabitants and Burghers to State Lands (Law of 13 July 1889). In Complete Collection of Laws of the Russian Empire (Polnoe sobranie zakonov Rossiiskoi imperii), 3rd series, Vol. 9, No. 6198, pp. 535–538. St. Petersburg: Printing House of the Second Section of His Imperial Majesty's Own Chancellery, 1891.

Anna Ioannovna. (1905). Decree of Empress Anna Ioannovna to Prince Shakhovskoi, ruler of Little Russia, on marriages of Little Russians (January 31, 1734). *Kievskaya Starina*, 91 (October 10).

Azbel, M. (1981). "Refusenik: Trapped in the Soviet Union", London: Hamish Hamilton.

Bakai, O. S., Bar'yakhtar, V. G., Gusynin, V. P., Zagorodny, A. G., Loktev, V. M., Peletminskii, S. V., Sitenko, Yu. O., Slyusarenko, Yu. V., Storizhko, V. Yu., & Shulga, M. F. (2020). *To the 90th birthday of Petro Ivanovych Fomin (1930–2011). Ukrainian Journal of Physics, 65*(8), 735–736. https://doi.org/10.15407/ujpe65.8.735

Bilenky S., Struminsky B. (n.d.). "Lviv National University". The *Internet Encyclopedia of Ukraine.* Available at: https://www.encyclopediaofukraine.com/display.asp?linkpath=pages%5CL%5CV%5CLvivNationalUniversity.htm (last consulted 3.05.2026).

Bogolyubov, A. N. (1996). N. N. Bogolyubov: Live, Creativity, Dubna: JINR

Bogolyubov, N. N., & Parasyuk, O. S. (1955). *O vychitatel'nom formalizme pri umnozhenii prichinnykh singulyarnykh funktsii* [On a subtraction formalism in the multiplication of causal singular functions]. *Doklady Akademii Nauk SSSR*, 100 (3) 429–432.

Bogoliubow, N. N., & Parasiuk, O. S. (1957). *Über die Multiplikation der Kausalfunktionen in der Quantentheorie der Felder* [On the multiplication of causal functions in quantum field theory]. *Acta Mathematica*, 97, 227–266. https://doi.org/10.1007/BF02392399

CERN (1992). *Relation with Non-Members States. Ukraine, 1992-1994*
CERN-ARCH-DSU-DO-02-2-17 https://cds.cern.ch/record/1953628?ln=en

CERN (2024). Jacques Haissinski (1935-2024). CERN Courier, July 2024. Available at: https://cerncourier.com/a/jacques-haissinski-1935-2024/ (last consulted 10.04.2026).

Chang, K. (2023). Stanley Deser, Whose Ideas on Gravity Help Explain the Universe, Dies at 92. The New York Times, 8-10 May 2023. Available at https://www.nytimes.com/2023/05/08/science/stanley-deser-dead.html (last consulted 28.04.2026).

Charpak, G. (2008). “Mémoires d'un déraciné, physicien, citoyen du monde”, Paris: Odile Jacob.

Chodorow (2005). Memorial resolution of Marvin Chodorow (1913-2005) https://web.archive.org/web/20111020145405/http://histsoc.stanford.edu/pdfmem/ChodorowM.pdf (last consulted 3.04.2026).

CMS Collaboration. (2025). *Measurement of light-by-light scattering and the Breit–Wheeler process, and search for axion-like particles in ultraperipheral PbPb collisions at* $\sqrt{s_{NN}}$ *= 5.02 TeV*. *Journal of High Energy Physics, 2025*(8), Article 6. https://doi.org/10.1007/JHEP08(2025)006.

Cockcroft, J.D. and Walton, E.T. (1932). Proc. Roy. Soc. A, 136, 619.

Constitution UkrSSR (1929). All-Ukrainian Central Executive Committee, *Constitution of the Ukrainian Socialist Soviet Republic*. Available at: https://zakon.rada.gov.ua/laws/show/n0001316-29#Text (last consulted 3.05.2026).

Cottin, M. (1979). “Monsieur Moïse HAÏSSINSKY au Laboratoire”, JOURNAL DE CHIMIE PHYSIQUE. 1979. 76, n° 7-8.

Cottrell, A. G., & Cottrell, L. S. (Eds.). (1995). *Times to remember: The life of Edward L. Ginzton*. Berkeley: Blackberry Creek Press.

Crease R. and Goldhaber A. (2012). Maurice Goldhaber 1911–2011, National Academy of Sciences. Available at http://biographicalmemoirs.org/pdfs/goldhaber-maurice.pdf (last consulted 3.04.2026).

Csörgő, T. (2025). In Memoriam László Jenkovszky, Low-X workshop. https://indico.cern.ch/event/1490864/contributions/6682446/attachments/3132867/5558039/2025-09-09-Csorgo-4-Jenkovszky-v3.pdf (last accessed 5.04.2026).

Dobrigkeit, C. (2026). Gleb Wataghin and his mission in Brazil, these proceedings.

Dzyuba, I.M. “RUSSIFICATION in the Ukrainian context”, Encyclopedia of Ukrainian History [Electronic resource]. Available at: https://www.history.org.ua/?termin=Rusyfikatsiia (last viewed: 29.03.2026).

ESU Baier Mykola Mykolaiovych / Yu. V. Piadyk, L. V. Bazhenov // Encyclopedia of Modern Ukraine [Online] / Eds.: I. M. Dziuba, A. I. Zhukovsky, M. H. Zhelezniak [et al.]; National Academy of Sciences of Ukraine, Shevchenko Scientific Society. – Kyiv: Institute of Encyclopedic Research, NASU, 2003. – Available at: https://esu.com.ua/article-38794 (last consulted 9.04.2026).

ESU Sorokin Pavlo Volodymyrovych / R. P. Slabospytskyi // Encyclopedia of Modern Ukraine [Online] / Eds.: I. M. Dziuba, A. I. Zhukovsky, M. H. Zhelezniak [et al.]; National Academy of Sciences of Ukraine, Shevchenko Scientific Society. – Kyiv: Institute of Encyclopedic Research, NASU, 2024. – Available at: https://esu.com.ua/article-883374 (last consulted 9.04.2026).

Fainberg, Ya. (2005). In Memory of Yakov Borisovich Fainberg (September 7, 1918–March 7, 2005). Plasma Physics Reports, Vol. 31, No. 10, 2005, pp. 891–892. Translated from Fizika Plazmy, Vol. 31, No. 10, 2005, pp. 959–960. Available at: https://link.springer.com/content/pdf/10.1134/1.2101977.pdf (last consulted 10.04.2026).

Gamow, G. (1928). Zur Quantentheorie des Atomkernes. Zeitschrift für Physik, 51, 204–212.

Gamow, G. (1970). "My world line", New York: The Viking Press.

Gerasimenko, V. (2026). Kyiv Origins of Modern Mathematical Physics, in these proceedings.

Gorenstein, M., Perepelytsya, S., Trubnikov, V. (2026). High Energy Density Physics at BITP: A Historical Overview, in these proceedings.

Grekov, V. (2026). Restoring the Narrative: Ways to Rethinking the Global Contribution of Ukrainian Physics, in these proceedings.

Grynyov, B., Hryn'ova, T. (2026). UKRAINE-CERN: the way to cooperation, in these proceedings.

Grynyov, B. V., Titov, M. P., & Stadnik, P. O. (2019). *Ukraine — CERN: The way to cooperation*. *Science and Innovation, 15*(5), 87–98. https://doi.org/10.15407/scine15.05.087

Haïssinski, M. (1957) La chimie nucléaire et ses applications. Paris: Masson.

Holovatch Yu., Honchar Y. and Krasnytska M. (2018), PHYSICS AND PHYSICISTS IN THE SHEVCHENKO SCIENTIFIC SOCIETY, Journal of physical studies, v. 22, No. 4 4003, https://doi.org/10.30970/jps.22.4003 (in Ukrainian); also in Leopolis Scientifica. Exact Sciences in Lviv until the middle of the 20th century. Lviv, 2021, pp. 273-302 (in English).

Hryn'ova, T., Grekov, V., and Spodenets, I. (2026). *Quantum of independence: How Ukraine is reclaiming its pioneering role in the history of modern physics*. ZN.UA. February 18. https://zn.ua/eng/quantum-of-independence-how-ukraine-is-reclaiming-its-pioneering-role-in-the-history-of-modern-physics.html.

Hull, M. (1998). Gregory Breit 1899-1981, Washington D.C.: National Academies Press. http://biographicalmemoirs.org/pdfs/breit-gregory.pdf (last consulted 3.04.2026).

Johnsen, K. (1977). Gersh Budker. Nature 270, 459. https://doi.org/10.1038/270459a0

Khramov, Yu. (1991). History of formation and development of physics schools in Ukraine, Kyiv: Phenix (in Russian).

Klyovanyy, L. (1928). “Geographical Atlas of Ukraine”, Kyiv: State Publishing House of Ukraine.

Kobchenko, K. (2020). Ukrainian emigration to Germany after WWII: Formation and Self-Representation of a Transnational Community (1945–1991), Euxeinos, Vol. 10, No. 30, https://gce.unisg.ch/fileadmin/user_upload/HSG_ROOT/Institut_GCE/Euxeinos/30/Euxeinos_30_71-89_Kobchenko.pdf (last accessed 5.04.2026).

Kozyrski, W. H., & Shenderovskyi, V. A. (2020). *To the history of theoretical researches at the Institute of Physics of NAS of Ukraine*. *Optoelectronics and Semiconductor Technique, 55*, 58–82. http://www.iop.kiev.ua/assets/files/2020/vtp_history.pdf

Kragh, H. (2025). Tachyons before tachyons: Lev Strum (1890–1936) and superluminal velocities. EPJ H 50, 9 (2025). https://doi.org/10.1140/epjh/s13129-025-00103-z

Lalli, R., & Navarro, J. (eds.) (2024). Globalizing Physics. Oxford: Oxford University Press.

Lavoisier, A. (1789). “Traité Élémentaire de Chimie”, Paris: Cuchet.

Lazur, V. (2021). "In Ukraine, I would have been sent to Siberia or shot": Lyubomyr Romankiv on his brilliant career in the USA and Apple, Radio Svoboda (in Ukrainian). https://web.archive.org/web/20210417201817/https://www.radiosvoboda.org/a/lubomyr-romankiv-ibm-90-rokiv/31203042.html (last access 4.04.2026).

Levchuk, L. (1992), The intra-atomic motion of bound electrons as a possible source of a systematic error in electron beam polarization measurements by means of a Moller polarimeter, KFTI 92-32, https://inis.iaea.org/records/58e6a-15s46 (last consulted 9.04.2026).

Levchuk, L. (2026). NSC KIPT in experimental particle physics: from 2 GeV electron Linac to CMS experiment at LHC. Available at: https://indico.cern.ch/event/1603776/contributions/6848846/ (last consulted 4.05.2026).

Linfield, H. S. (1928). The American Jewish Year Book, Vol. 31 (October 5, 1929, to September 22, 1930 / 5690), pp. 299-334, https://www.jstor.org/stable/pdf/23601425.pdf (last accessed 4.04.2026).

Lock, W.O. (1975). A history of the collaboration between the European Organization for Nuclear Research (CERN) and the Joint Institute for Nuclear Research (JINR), and with Soviet research institutes in the USSR 1955-1970, CERN-75-07, https://cds.cern.ch/record/186009 (last accessed 3.04.2026).

Lucie Str. (1944). “I married a Russian : Letters from Kharkov”, London : George Allen & Unwin Ltd.

Martynov, E., Nicolescu, B. (2019). Odderon effects in the differential cross-sections at Tevatron and LHC energies. Eur. Phys. J. C 79, 461. https://doi.org/10.1140/epjc/s10052-019-6954-6 (last consulted 9.04.2026).

Neklyudov, I. M., et al. (2004). *In memory of Mykhailo Petrovych Rekalo*. *Problems of Atomic Science and Technology*, (4), 155–156. https://vant.kipt.kharkov.ua/ARTICLE/VANT_2004_4/article_2004_4_155.pdf

Novosyadlyj, B. (2026). George Gamow: from beta decay to cosmological nucleosynthesis, in these proceedings.

Nurmagambetov, A. J. (2026). Prof. D. V. Volkov at CERN, in these proceedings.

Oxford English Dictionary (n.d.), Russian (noun and adjective) https://www.oed.com/dictionary/russian_n (last viewed 12.05.2026).

Panofsky, W. K. H. (2007). *Panofsky on physics, politics, and peace: Pief remembers*. Springer. https://doi.org/10.1007/978-0-387-69732-1.

Petrovsky-Shtern, Y., and Polonsky, A. (Eds.). (2014). *POLIN: Studies in Polish Jewry* (Vol. 26, *Jews and Ukrainians*). The Littman Library of Jewish Civilization. https://doi.org/10.2307/j.ctv2vr8vjf.

Plokhy, S. (2010). “The origins of the Slavic Nations”. Cambridge: Cambridge University Press.

Plokhy, S. (2015). “The Gates of Europe”. London: Allen Lane.

PT (1961). Physics Today January 1961, https://aip.brightspotcdn.com/PTO.v14.i1.98_3.online.pdf (last viewed 2.04.2026).

Pulyui, I. (1996). Collection of Works. Edited by V. Shenderovskyi, Kyiv: Rada (in Ukrainian).

Ranyuk,Y. (2006), Laboratory N. 1: history of nuclear physics in Ukraine, Kharkiv: ACTA.

RASA, https://rasa-usa.org/en/ (last viewed: 2.04.2026).

Rovenchak O. and Volodko V. (2018), “Comparative Analysis of the Identity and Practices of Post-war Representatives and the Contemporary Wave of Ukrainian Immigrants to the United States”, Annales Universitatis Mariae Curie-Sklodowska Lublin - Polonia, vol. XLIII, I, p. 33. Available at https://journals.umcs.pl/i/article/download/6541/5801 (last viewed: 20.03.2026).

Rutherford (1911). “The Scattering of α and β Particles by Matter and the Structure of the Atom.” *Philosophical Magazine*, Series 6, **21** (125), 669–688.

Schenderovskyi, V. (2026). The celebrated legacy of the eminent 20th-century physicist Dmytro Ivanenko, in these proceedings.

Scherbak, O. (2012). Nathan Rozen in Ukraine: unknown pages of life and work, Visnyk NTU KhPI, 42 (948) p. 179. (in Ukrainian) https://library.kpi.kharkov.ua/files/Vestniki/2012_42.pdf.

Shepelev, A. and Larbalestier, D. (2011). The discovery of type II superconductors, CERN Courier, October 2011. Available at https://cerncourier.com/a/the-discovery-of-type-ii-superconductors/ (last consulted 10.04.2026).

Shiltsev, V. (2025). Post 1980s Diaspora of Russian Particle Physicists. 4th international symposium on history of particle physics. CERN. 10-13 November. https://indico.cern.ch/event/1480892/contributions/6386322/.

Shifman, M., Vainshtein, A., Shuryak E., and Zelevinsky V. (2025). Iosif Khriplovich 1937–2024 CERN Courier, May 2025. https://cerncourier.com/a/iosif-khriplovich-1937-2024/ (last consulted 9.04.2026).

Sidletskiy, O. (2025). Ukrainian crystals for the Large Hadron Collider: when science is stronger than war. Mirror of the Week, October 22. Available at: https://zn.ua/eng/ukrainian-crystals-for-the-large-hadron-collider-when-science-is-stronger-than-war.html (last consulted 4.05.2026).

Sinelnikov, C., Leipunsky, A., Walther, A., and Latischev, G. (1932). The disintegration of lithium by high velocity protons. Physikalische Zeitschrift der Sowjetunion, 2(3), 285.

Snyder, T. (2003). “The reconstruction of nations : Poland, Ukraine, Lithuania, Belarus, 1569-1999”. New Haven & London: Yale University Press.

Snyder, T. (2015). “Bloodlands”, London: Vintage.

SSCU (2001). State Statistics Committee of Ukraine “All-Ukrainian population census”. Available at: https://web.archive.org/web/20110811204954/https://2001.ukrcensus.gov.ua/results/general/nationality/ (last viewed: 30.03.2026).

Stober, D. (2009). Physicist: Universes pop up ad infinitum, Stanford report, April 1st, https://news.stanford.edu/news/2009/april1/hofstadter-lecture-tufts-cosmologist-alexander-vilenkin-040109.html (last viewed: 3.04.2026).

Strikha M.V. (2024). Ukrainian language and physics: a brief historical retrospective (to the 150th anniversary of the Shevchenko scientific society). Ukr. J. Phys. 69, No. 2, 124. https://doi.org/10.15407/ujpe69.2.124

Swartz, M. (2025). https://indico.cern.ch/event/1480892/contributions/6377879/attachments/3170044/5635322/SLD_Electroweak_Physics_M_Swartz.pdf (last consulted 9.04.2026).

Sytov, A., et al. (2022). *First design of a crystal-based extraction of 6 GeV electrons for the DESY II Booster Synchrotron*. European Physical Journal C, 82 (3), Article 197. https://doi.org/10.1140/epjc/s10052-022-10115-4

Tan’shyna, A. (2019). Titans of Ukraine, Ukr. J. Phys. 2019. Vol. 64, No. 1 https://ujp.bitp.kiev.ua/index.php/ujp/article/view/2019294/1320 (consulted 31.03.2026).

Tarasov, V. (2026). ISMA's CsI scintillators in experiments on high-energy physics and astrophysics, in these proceedings.

Teryaeva, N. (2019). “Academician Khalatnikov turns 100!”, Otkrytanya Dubna, https://web.archive.org/web/20201031153708/https://open-dubna.ru/obshchestvo/7296-akademiku-khalatnikovu-sto-let (last viewed 3.04.2006).

Thomson, J. J. (1897). “Cathode Rays.” *Philosophical Magazine*, Series 5, **44** (269), 293–316.

Timoshenko, S. (1968). As I remember Princeton, N.J.: Van Nostrand.

Timoshyn D., Hryn’ova T., Kostiuk I. (2025), Ukrainian contribution to particle physics: historical perspective and prospects, PoS ICHEP2024 1197, https://arxiv.org/abs/2501.03878

Tkachov, F. V. (2009). *A contribution to the history of quarks: Boris Struminsky's 1965 JINR publication* (arXiv:0904.0343). arXiv. https://doi.org/10.48550/arXiv.0904.0343

Tupitsyna, I. (2026). PWO Scintillator: How the Story Began, in these proceedings.

USSR (1959), All-Union Population Census. https://web.archive.org/web/20241206162837/https://www.demoscope.ru/weekly/ssp/sng_nac_59.php?reg=2 (accessed 5.04.2026).

USSR (1970), All-Union Population Census. https://web.archive.org/web/20251212092033/https://www.demoscope.ru/weekly/ssp/sng_nac_70.php?reg=2 (accessed 5.04.2026).

VRU (1992). “About the state border of Ukraine (Про державний кордон України)”, Information of the Verkhovna Rada of Ukraine (Відомості Верховної Ради України), 1992, № 2, article 5. Available at https://zakon.rada.gov.ua/laws/show/1777-12#Text (accessed March 29, 2026).

Weissberg-Cybulski, A. (1951). The Accused. Simon and Schuster.

Weisskopf, V. (1972). Physics in the Twentieth Century: Selected Essays, Cambridge (USA) and London (England): The MIT Press.

Weisskopf, V. (1991). The joy of insight, BasicBooks.

Yanovsky, F. (2016). Glimpses of Early Radar Developments in Ukraine and the Former Soviet Union, Radio Science Bulletin 358, pp. 35-68.

Yarba, V. (2019). Here and There, or There and Here A Physicist's Recollection: Science in the midst of political turmoil in the USSR and USA (Russian edition published Dubna: JINR in 2019, English edition published independently in 2022).

Zagorodnii, A.G., et al. (2022). *To the 100th anniversary of the outstanding scientist Ostap Stepanovych Parasyuk (20.12.1921–22.11.2007). Ukrains'kyi Matematychnyi Zhurnal*, 74(2), 279–283. https://doi.org/10.37863/umzh.v74i2.6873

Zavorotna, N. (2020). In Scholars in Exile: The Ukrainian Intellectual World in Interwar Czechoslovakia Pages: 53–71 University of Toronto Press, https://utppublishing.com/doi/10.3138/9781487504458.004 (last accessed 4.04.2026).

Zhuk, S. (2024). The KGB, Russian Academic Imperialism, Ukraine, and Western Academia 1946-2024, Lanham: Lexington Books.

Zhuk, S. (2026). *Academic exchanges and technological/industrial espionage of the GRU/KGB in the West*, in these proceedings.